\documentclass[lettersize,journal]{IEEEtran}
\usepackage{amsmath,amssymb,amsfonts} 
\usepackage{booktabs} 
\usepackage{algorithm}
\usepackage{array}
\usepackage[caption=false,font=normalsize,labelfont=sf,textfont=sf]{subfig}
\usepackage{textcomp}
\usepackage{stfloats}
\usepackage{float}
\usepackage{threeparttable} 
\usepackage{multirow}  
\usepackage{url}
\usepackage{verbatim}
\usepackage{graphicx}
\usepackage{cite}
\usepackage[utf8]{inputenc} 
\usepackage{color} 
\usepackage{bbding} 
\usepackage{makecell} 
\usepackage{epstopdf} 
\usepackage[noend]{algpseudocode} 
\algdef{SE}[DOWHILE]{Do}{doWhile}{\algorithmicdo}[1]{\algorithmicwhile\ #1}
\begin{document}

\title{Exploring Siamese Networks in \\ Self-Supervised Fast MRI Reconstruction}

\author{Liyan Sun, Shaocong Yu, Chi Zhang and Xinghao Ding

\thanks{This work was supported in part by the National Natural Science Foundation of China under Grants 82172033, U19B2031, 61971369, 52105126,  U1805261 and 22161142024. (The co-first authors, Liyan Sun and Shaocong Yu, contributed equally to this work.) (Corresponding author: Xinghao Ding (e-mail: dxh@xmu.edu.cn).)}

\thanks{L. Sun is with Stanford University School of Medicine, Stanford University, Stanford CA 94035, USA.}

\thanks{S. Yu and C. Zhang are with school of Electronic Science and Engineering, Xiamen University, Xiamen 361005, China.}

\thanks{X. Ding is with the School of Informatics, Xiamen University, Xiamen 361005, China and Institute of Artificial Intelligence, Xiamen University, Xiamen 361005, China
.}}

\markboth{Journal of \LaTeX\ Class Files,~Vol.~14, No.~8, February~2023}%
{Shell \MakeLowercase{\textit{et al.}}: Exploring Siamese Networks in Self-Supervised Fast MRI Reconstruction}

\IEEEpubid{0000--0000/00\$00.00~\copyright~2023 IEEE}

\maketitle

\begin{abstract}
Reconstructing MR images using deep neural networks from undersampled k-space data without using fully sampled training references offers significant value in practice, which is a self-supervised regression problem calling for effective prior knowledge and supervision. The Siamese architectures are motivated by the definition ``invariance'' and shows promising results in unsupervised visual representative learning. Building homologous transformed images and avioding trivial solutions are two major challenges in Siamese-based self-supervised model. In this work, we explore Siamese architecture for MRI reconstruction in a self-supervised training fashion called SiamRecon. We show the proposed approach mimics an expectation maximization algorithm. The alternative optimization provide effective supervision signal and avoid collapse. The proposed SiamRecon achieves the state-of-the-art reconstruction accuracy in the field of self-supervised learning on both single-coil brain MRI and multi-coil knee MRI.
\end{abstract}

\begin{IEEEkeywords}
MRI Reconstruction, Self-Supervised Learning, Siamese Networks, Deep Learning.
\end{IEEEkeywords}

\section{Introduction}
\IEEEPARstart{M}{agnetic} Resonance Imaging (MRI) is a widely used medical imaging technique for its advantages such as multi-parametric information, high resolution in soft tissues and low radiation. However, the long scanning time in MRI examinations is disadvantageous, which can cause patients' discomfort and motion artifacts. Therefore, numerous methods were proposed to accelerate MRI \cite{lustig2007sparse,pruessmann1999sense,griswold2002generalized,sun2016deep}. One stream of the research in fast MRI explores the usage of compressed sensing theory, referred to as Compressed Sensing MRI (CS-MRI) \cite{lustig2007sparse,sun2016deep,song2019coupled}. K-space measurements are acquired partially and randomly while the low-frequency bands are sampled densely to capture the major energy of this signal in CS-MRI. Then certain prior knowledge like total variation \cite{yao2018efficient}, wavelet sparsity \cite{qu2012undersampled} or dictionary learning \cite{song2019coupled,ravishankar2010mr} is  imposed on the nonlinear reconstruction of an undersampled MR image to approximate to its latent full-sampled one. Under a multi-coil setup, CS-MRI can be incorporated into parallel MRI to further accelerate imaging \cite{liang2009accelerating}.

Deep neural networks have been used for both single-coil and multi-coil MRI reconstruction successfully \cite{sun2016deep,schlemper2017deep,duan2019vs,arvinte2021deep}. Compared with conventional methods, deep neural networks leverage rich historical and external data to extract image representations for different visual tasks. However, the current major study of deep neural networks for CS-MRI suppose massive fully sampled MRI training references are available and indispensable. This prerequisite poses major problems in the following three folds. First, full acquistion in k-space could produce poor-quality ``golden standard'' training references. For example, the slow full data acquisition corrupts the image with motion artifacts in high-resolution dynamic cardiac \cite{usman2013motion} or abdominal imaging \cite{zaitsev2015motion}. Second, T2$^\star$ relaxation tends to decay rapidly in some sequences like echo planar imaging \cite{clifford2021artificial}, making time-consuming full acquisition impractical. Third, collecting fully sampled training references from other sites could introduce domain shift. Developing self-supervised deep learning models in which training samples are only partially acquired k-space measurements themselves is promising to overcome the aforementioned difficulties. Yet the self-supervised fast MRI reconstruction is also an open and challenging problem.

Learning end-to-end mapping from undersampled MRI inputs to their fully sampled MRI references is infeasible in self-supervised learning, so regularization or supervision based on observations themselves is critical to enable effective training.
To exploit the corrputed observations, multiple training strategies for deep neural networks were proposed for self-supervised MR image recovery \cite{yaman2020self,hu2021self}. An approach for Self Supervised learning via Data Undersampling (SSDU) partitions the undersampled k-space measurements into a training component and a loss component. The reconstruction network takes as input the training component, and the k-space values of the network output should be identical to the loss components on sampled positions serving as a supervision signal. In a recently proposed dual networks for self-supervised CS-MRI \cite{hu2021self}, the undersampled k-space data is partitioned into k-space subsets twice, which are input to two independent deep networks. Besides the data undersampling strategy developed in SSDU, a similarity penalty is imposed based on the assumption that the dual networks should offer identical estimations on the unscanned k-space positions. The concept of homology is explored in the dual networks, but the model may be at risk of collapsing if the dual networks ``collaborate'' without stable supervision and get stuck in local optima.

\IEEEpubidadjcol
Based on above analysis, proposing an effective training strategy to exploit obvervations with stable and reasonable supervision is the remedy for the exisiting difficulty in self-supervised CS-MRI. Recently, a Simple Siamese (SimSiam) architecture \cite{chen2021exploring} was successfully applied in unsupervised learning. Motivated by the prior about image invariance, a Siamese architecture takes two augmented views from one image, then the model is forced to generate consistent features with a stop-gradient strategy. The primary SimSiam model is used for feature learning in high-level visual tasks and demonstrates promising results in avoiding collapse. Inspired by this work, we explore the Siamese networks in self-supervised CS-MRI.

In this paper, we propose a Siamese architecture called SiamRecon for self-supervised learning in fast MRI reconstruction. The undersampled k-space measurements offer one view naturally of the latent fully sampled MRI. We augment other views using a k-space resampling strategy. A homology prior is applied by imposing the consistency between those views. We first formulate the SiamRecon architecture from the perspective of Expectation Maximization (EM). In this way, the model training is converted to the alternations between MR image reconstruction given network parameters and optimization of network parameters given temporary reconstructed MR images. Such a alternate optimization enables stable supervision signal  with gradient stop operation. The experimental results on both single-coil brain MRI and multi-coil knee MRI demonstrates the state-of-the-art reconstruction accuracy among compared self-supervised MRI methods.

\section{Related Works}
\subsection{Fast MRI}
Compressed sensing enables fast MRI reconstruction by partially sampling the k-space data and then perform nonlinear reconstruction by prior knowledge. Lustig et.al. proposed a SparseMRI method using both the total variation regularization and wavelet coefficient sparsity \cite{lustig2007sparse}. Non-local prior was combined with sparse prior for CS-MRI reconstruction \cite{qu2014magnetic}. Dictionary learning methods enables adaptive sparse representative learning  in situ \cite{song2019coupled,ravishankar2010mr,huang2014bayesian}.

Deep learning model is first introduced into CS-MRI with a convolutional neural network using end-to-end mapping from a zero-filled MR image to its fully-sampled one. Then a deep cascaded convolutional neural networks \cite{schlemper2017deep} was proposed for dynamic MRI reconstruction, where a data consistency term can be embedded as differentiable network layers. The model-driven CS-MRI method were also used for parallel MRI reconstruction \cite{aggarwal2018modl}. Adversarial learning was used to render the reconstructed MR images more realisitc \cite{quan2018compressed,shaul2020subsampled}. Recently, undersampling trajectory was optimized jointly with reconstruction networks targeting on the entire image \cite{bahadir2020deep} or region of interests \cite{sun2021fast}. The deep learning methods achieved significant improvements in reconstruction accuracy on the condition of plentiful fully-sampled  training references. However, fully-sampling can not guarantee high-quality reconstructed MR image with fast moving imaging regions or specific imaging sequences.

In order to effectively train a reconstruction network without fully sampled training references, Yaman et.al. proposed the SSDU method \cite{yaman2020self} to decompose undersampled k-space observations to create input k-space data and supervision signal based on the concept of cross validation. A dual network further developed a similarity loss based on the assumption that two networks takes different training components of one original partial k-space observations should produce consistent results \cite{hu2021self}. However, developing stable supervision and avoiding trivial solution in such invariance regularization still remain to be addressed.

\subsection{Self-Supervised Learning}
Self-supervised learning aims at digging up regularity and patterns within unlabeled/ill-labled data \cite{jing2020self} via pretexts like image colorization \cite{larsson2017colorization}, image inpainting \cite{pathak2016context} or rotation prediction \cite{komodakis2018unsupervised}. The pretexts leverage the prior knowledge of images like appearance or physics to extract semantic features. Constrastive learning further composes positive/negative examples and group positive ones while diverse negative ones \cite{ye2019unsupervised,bachman2019learning,tian2020contrastive,he2020momentum,chen2020simple}. In a MOmentum COntrast (MOCO) model \cite{he2020momentum}, a dual encoder network was developed with a queue of negative data samples to build dynamic dictionaries. The SimCLR \cite{chen2020simple} spared the need for memory bank by composing data augmentations, using nonlinear head and applying larger batch size. The BYOL \cite{grill2020bootstrap} further spared the usage of negative examples and avoided collapse via a momentum encoder. Based on the Siamese architecture to model invariance, a SimSiam method directly maximaizes the similarity of one image's two views with neither negative samples nor memory bank. A Barlow twins avoids collapse by imposing the cross-correlation matrix to be an identity matrix between the outputs from Siamese networks \cite{zbontar2021barlow}.

\section{Method}
\subsection{Reconstruction Networks}
For the single-coil compressive sensing MRI, the objective function to reconstruct a vectorized MR image $x \in {\mathbb{C}^{N \times 1}}$ is written as
\begin{equation}\label{eq1}
\arg \mathop {\min }\limits_x \left\| {{y_\Omega } - {F_\Omega }x} \right\|_2^2 + \lambda \mathcal{R}\left( x \right),
\end{equation}
where subscript $\Omega$ represents the sampled positions in k-space and $y_\Omega$ is the partially acquired k-space measurements. The partial Fourier matrix $F_\Omega \in {\mathbb{C}^{M \times N}} \ \left( {M < N} \right)$ performs full Fourier transform of the underlying ground-truth image and then only keep the sampled k-space positions. The first term ensures the consistency between the reconstructed MR image and k-space observations. The second term is the prior regularizing the ill-posed nature of CS-MRI.  We adopt the ISTA-Net \cite{zhang2018ista} as the model-based reconstruction network for single-coil CS-MRI. The Eq. \ref{eq1} is further formulated as
\begin{equation}\label{eq2}
\arg \mathop {\min }\limits_x \frac{1}{2}\left\| {{y_\Omega } - {F_\Omega }x} \right\|_2^2 + \lambda {\left\| {g\left( x \right)} \right\|_1},
\end{equation}
where a $\mathcal{L}_1$ sparsity regularization is imposed on the adaptively transformed coefficients ${g\left( x \right)}$ with regularization parameter $\lambda$. The optimization of Eq. \ref{eq2} can be iteratively performed by following two steps:
\begin{equation}\label{eq3}
\begin{array}{l}
{r^{\left( j \right)}} = {x^{\left( {j - 1} \right)}} - \rho F_\Omega ^H\left( {{F_\Omega }{x^{\left( {j - 1} \right)}} - {y_\Omega }} \right)\\
{x^{\left( j \right)}} = \widetilde g \left( {soft\left( {g \left( {{r^{\left( j \right)}}} \right),\theta } \right)} \right),
\end{array}
\end{equation}
where $j$ indicates the iteration index, and $\rho$ is the step size. The adaptive transform ${g\left( \cdot \right)}$ and its left inverse ${\widetilde g\left( \cdot \right)}$ are achieved using convolution blocks. The soft-thresholding operator is denoted as $soft\left( \cdot \right)$ using the threshold $\theta$. The iteration is unfolded and all the learnable parameters can be optimized using error backpropagation in an end-to-end manner.

\begin{figure}[t]
\centerline{\includegraphics[width=0.85\columnwidth]{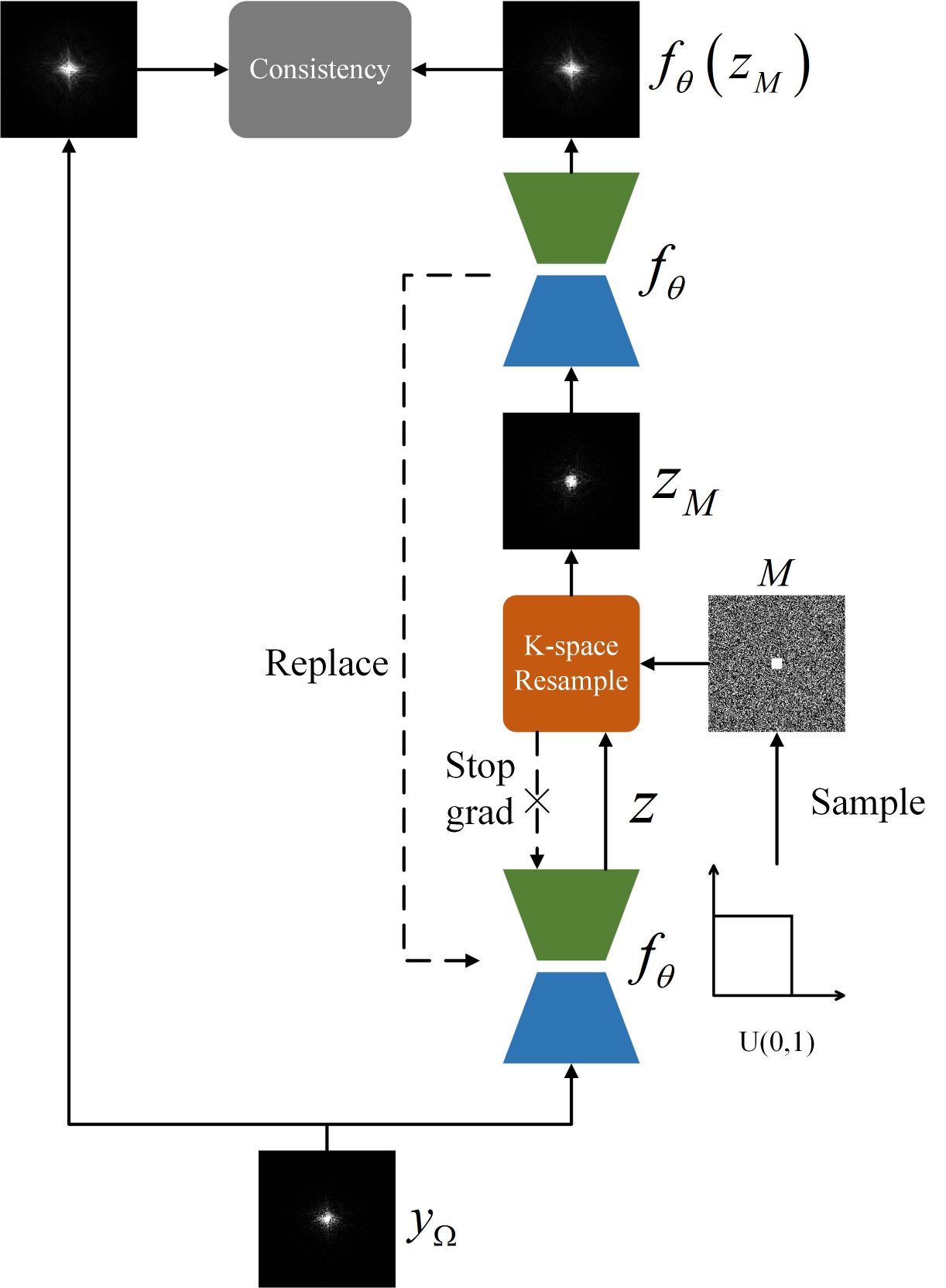}}
\caption{The architecture of the proposed SiamRecon.}
\label{fig1}
\end{figure}

For the parallel compressive MRI, supposing $n_c$ coils are deployed, the Eq. \ref{eq1} is rewritten as
\begin{equation}\label{eq4}
\arg \mathop {\min }\limits_x \sum\limits_{i = 1}^{{n_c}} {\left\| {{y_{\Omega ,i}} - {F_\Omega }{S_i}x} \right\|_2^2}  + \lambda R\left( x \right),
\end{equation}
where $y_{\Omega,i}$ is the partially acquired k-space measurements on the $i$-th coil. The $i$-th coil sensitivity map $S_i \in {C^{N \times N}}$ is a diagonal matrix which can be precomputed using E-SPIRiT \cite{uecker2014espirit} algorithm. The model-based VS-Net method \cite{duan2019vs} admits a closed-form solution to the optimization of the first term in Eq. \ref{eq4} and provide efficient computation. The variable splitting algorithm is unfolded into a end-to-end learnable deep neural networks to compose a VS-Net architecture. Each iteration has three blocks including a denoiser block, a data consistency block and a weighted average block. The detailed description of the ISTA-Net \cite{zhang2018ista} and VS-Net \cite{duan2019vs} can be found in related works. It is noted that both the two models assume available fully sampled training references in their original papers, and we use them as base network architectures.

\subsection{SiamRecon}
\subsubsection{Model Architecture}
Inspired by the Simple Siamese network architecture, we propose a SiamRecon whose architecture is shown in Fig. \ref{fig1}. In the proposed SiamRecon, we can treat the unknown fully sampled MRI as a latent variable, its one view can be naturally derived as the acquired partially sampled MRI measurements $y_{\Omega}$. Motivated by a homology prior, we augment other views.

First, we reconstruct a MRI using a parameterized network $f_{\theta}$ as an intermediate variable $z = f_{\theta} \left( y_{\Omega} \right)$. Note $z$ is reconstructed MRI in Fourier domain. Then we develop a k-space resampling to augment another view $z_M=Mz$ using a binary mask $M$. The binary mask is randomly drawn from a uniform distribution ${\mathcal U} \left( 0,1 \right)$. The central regions are densely sampled to preserve most energy of the intermediate reconstruction. In our paper, we use 1D line-based or 2D point-based resampling mask depending on the undersampling pattern. The manually augmented view $z_M$ is also derived from the latent fully sampled MRI. We again use the network $f_{\theta}$ to reconstruct MRI from the view $z_M$. Then a consistency penalty can be imposed on the two views to train the network $f_{\theta}$. As shown in Figure \ref{fig1}, we apply stop gradient operation to fix the intermediate variable $z$, which helps stabilize model training. With $f_{\theta} \left( z_M \right)$ that approximates the observation $y_{\Omega}$, we update the Siamese network and recompute the intermediate variable $z$ to continue model training. In the next section, we discuss the link between the proposed SiamRecon with a EM-like algorithm.

\subsubsection{EM-Like Formulation}
In this section, we start by formulating the our approach from the perspective of Maximum Likelihood Estimation (MLE). The forward model is definite and denoted as $y_{\Omega}=F_{\Omega} x$. We parameterize the inversion model as $f_{\theta} \left( {y_{\Omega}} \right)$ expected to approximate the latent fully sampled MRI $x$. The MLE aims at optimizing the following objective function,
\begin{equation}\label{eq5}
\begin{aligned}
\widehat \theta  &= \arg \mathop {\max }\limits_\theta  \left[ {\log p\left( {{y_\Omega } |\theta} \right)} \right]\\
 &= \arg \mathop {\max }\limits_\theta  \left[ {\log \left( {\int_x {p\left( {{y_\Omega }|\theta ,x} \right)p\left( x \right)dx} } \right)} \right],
\end{aligned}
\end{equation}
where we maximize the probability of the observed partial k-space measurements $y_{\Omega}$ given the network parameter $\theta$. Since no fully sampled image $x$ can be used as training reference, we introduce it as a latent variable. The marginal integration of $x$ in the first term of Eq. \ref{eq5} is intractable. Accordingly, a EM approach is used to optimize model parameters in Eq. \ref{eq5}.

The EM algorithm involves the alternations between solving two sub-problems denoted as the E-step and the M-step until convergence:
\begin{equation}\label{eq6}
\begin{aligned}
&\text{E-step:}\ Q\left( x \right) := p\left( {x|{y_\Omega },\theta } \right)\\
&\text{M-step:}\ {\widehat \theta}{\rm{ }} = \arg \mathop {\max }\limits_\theta  \left\{ {\mathbb{E}{_{x \sim Q\left( x \right)}}\left[ {\log p\left( {{y_\Omega }|x, \theta} \right)} \right]} \right\}.
\end{aligned}
\end{equation}
Specifically, the E-step calculates the probability of the fully sampled MR image given the scanned partial k-space measurements $y_\Omega$ and the temporary estimation of the model parameter $\theta$. The M-step computes the expected value of log likelihood given network parameter $\theta$ to be optimized and the temporary latent variable $x$. The EM optimization has a nice property in convergence, but a global optima cannot be guaranteed. Hence a good initialization of the inversion model $\theta$ is indispensable.

%

\subsubsection{EM-Like Solution}
Based on the EM formulation, we explains the SiamRecon as follows. The E-step in the $t$-th iteration is computed by reconstructing MRI using $y_ \Omega$ and $\theta$ estimated in the $\left(t-1\right)$-th iteration:
\begin{equation}\label{eq7}
{z^{\left( t \right)}} = {f_{{\theta ^{\left( {t - 1} \right)}}}}\left( {{y_\Omega }} \right),
\end{equation}
where the network outputs Fourier coefficients of the reconstructed MR image ${z^{\left( t \right)}}$. In a model-based reconstruction network like ISTA-Net and VS-Net, a data consistency layer ensures the consistency between the ${z^{\left( t \right)}}$ and the $y_{\Omega}$ on sampled trajectory. As a result, we will have a trivial loss function if we apply a regular MLE optimization in the M-step
\begin{equation}\label{eq71}
{\theta ^{\left( t \right)}} = \arg \mathop {\min }\limits_\theta  {\rm{Dis}}\left( {{f_\theta }\left( {{z^{\left( t \right)}}} \right){|_\Omega },{y_\Omega }} \right),
\end{equation}
where $\rm{Dis \left( {\cdot} \right)}$ denotes certain measure of distance like $\mathcal{L}_1$ or $\mathcal{L}_2$, and the network $f_{\theta}$ simply needs to be an identical mapping to perfectly fit this constraint.

Inspired by the homology prior, we avoid trivial solution by developing an alternative loss function in the M-step
\begin{equation}\label{eq8}
{\theta ^{\left( t \right)}} = \arg \mathop {\min }\limits_\theta  {\rm{Dis}}\left( {{f_\theta }\left( {z_M^{\left( t \right)}} \right){|_\Omega },{y_\Omega }} \right),
\end{equation}
where $M$ is the aforementioned undersampling mask for k-space resampling, and $z_M^{\left( t \right)}$ is the resampled k-space data. The optimization of network parameter $\theta$ utilize the homology assumption that the undersampled k-space measurements $y_{\Omega}$ as one view should be consistent with manually augmented other views ${z_M^{\left( t \right)}}$ in the $t$-th iteration as the network $f_{\theta}$ serves as the predictor used in SimSiam model \cite{chen2021exploring}. Our maximization step provides stable supervision $y_{\Omega}$ for updating network with temporarily fixed $z$, which is a stop-gradient operation in helping avoid collapse.

With proper network parameter initialization, we have the initial reconstructed image by solving Eq. \ref{eq7}. Then this reconstructed MR image is resampled in Fourier domain as the input to another siamese network supervised by the partial k-space measurements, which is our homology penalty. We optimize this objective function in regard to network parameter $\theta$ until convergence. Then we update the the Siamese networks and have the reconstructed MR image again in Eq. \ref{eq7} to continue the loop. The whole training strategy of our proposed SiamRecon is described in Alg. \ref{alg}, where we use one training sample as illustration for simplicity and clarity. When network training ends, only network $f_{\theta}$ is needed in inference.

\begin{algorithm}[t]
\caption{Training Strategy of SiamRecon}
\label{alg}
{\bf Input:}
Undersampled MRI training samples ${y_\Omega }$. \\
{\bf Initialization:}
Train a ISTA-Net according to \cite{yaman2020self} to have a initialized network parameters ${\theta ^{\left( 0 \right)}}$. Set a threshold $\tau$.
\begin{algorithmic}[1]

\For{$t=0$ to $T-1$}
\State Estimate a full-sampled MRI ${z^{\left( t \right)}}$ using $f_{\theta^{\left( t \right)}}$;
\State Resample ${z^{\left( t \right)}}$ using mask $M$ to have ${z_M^{\left( t \right)}}$;
\State let $k=0$;
\State ${\theta_k ^{\left( t \right)}} ={\theta ^{\left( t \right)}}$;
\Do
\State Compute the gradient $\nabla \theta _k^{\left( t \right)}$ from Eq. \ref{eq8};
\State $\theta _{k+1}^{\left( t \right)} = {\theta_k ^{\left( t \right)}} - \alpha \nabla \theta _k^{\left( t \right)}$;
\State $k = k + 1$;
\doWhile
{{${\left\| {{f_{\theta _k^{\left( t \right)}}}\left( {z_{\Omega}^{\left( t \right)}} \right) - {f_{\theta _{k - 1}^{\left( t \right)}}}\left( {z_{\Omega}^{\left( t \right)}} \right)} \right\|_2} \geq \tau$}}
\State Update $\theta$ by ${\theta ^{\left( t+1 \right)}} = \theta _k^{\left( t \right)}$.
\EndFor

\end{algorithmic}
{\bf Output:}
Trained ISTA-Net with parameter ${\theta ^{\left( T \right)}}$.
\end{algorithm}

\section{Implementation}
The EM algorithm is susceptible to the parameter initialization: poor starting point directs the reconstruction model to bad local optima, especially in self-supervised learning where no clean image is available to provide full supervision. In this regard, we use the SSDU method \cite{yaman2020self} to pretrain a reconstruction network as initialization. Similar to \cite{yaman2020self,hu2021self}, we use ISTA-Net \cite{zhang2018ista} and VS-Net \cite{yaman2020self} as the deep network architecture. For the generation of resampling mask, we draw a sample from a uniform distribution with the central 20$\times$20 regions fully sampled. Here we use a normalized hybrid ${\mathcal L}_{1}$-${\mathcal L}_{2}$ loss function as the distance measure to impose consistency \cite{yaman2020self}.
\begin{equation}
{{\cal L}_{{\rm{hyb}}}}\left( {u,{y_\Omega }} \right) = \frac{{{{\left\| {u - {y_\Omega }} \right\|}_2}}}{{{{\left\| y_\Omega \right\|}_2}}} + \frac{{{{\left\| {u - {y_\Omega }} \right\|}_1}}}{{{{\left\| y_\Omega\right\|}_1}}},
\end{equation}
where $u$ represents the output reconstructed MRI ${{f_\theta }\left( {{z_M^{\left( t \right)}}} \right){|_\Omega }}$ in Fourier domain.

In our implementation of SiamRecon with ISTA-Net, we use Adam \cite{kingma2014adam} as optimizer with learning rate set to 1e-5. The ISTA-Net is unfolded to 9 phases. And the batchsize is set to 1. We adopt the same parameter setting in the implementation of SiamRecon with VS-Net, except the VS-Net architecture is unfolded to 5 cascaded blocks. The network parameters are optimized with 20 epochs. Our approach is implemented using Pytorch on a Nvidia GeForce GTX 1080 Ti.

\section{Experimental Results}
\label{sec:guidelines}

\subsection{Datasets}

\subsubsection{IXI Dataset}
We use the IXI Dataset for numeric experiments on single-coil brain MRI, which was acquired and collected from three hospitals in London. Multi-parametric MRI including T1, T2 and PD weighted images, MRA images 15-direction diffusion-weighted images. Here we use T1 weighted modality due to its ability to represent stucture within brains. The volume data is converted into 2D slices. The matrix size of each slice is 256$\times$256. The training, validation and testing dataset include 850, 250 and 250 slices, respectively. Here we use both 1D line-based Cartesian undersampling pattern and 2D point-based random undersampling pattern.

\subsubsection{FastMRI Knee Dataset}
We use the clinical multi-coil knee MRI publicized in \cite{knoll2020fastmri}. The dataset contains 20 imaging subjects, and each subject has total five image acquisition protocols including coronal PD, coronal saturated PD, axial fat-saturated T2, sagittal fat-saturated T2 and sagittal PD. We use coronal PD modality throughout this work. An off-the-shelf 15-element coil produces scans of approximately 40 slices. The provided coil sensitivity maps were precomputed according to the central 24$\times$24 fully sampled k-space block using BART toolbox \cite{uecker2015berkeley}. The golden standard is defined as the coil-sensitivity combined fully sampled reconstruction. We split the 20 imaging subjects into a training dataset of 10 subjects and a testing dataset of 10 subjects. Here we use 1D line-based Cartesian undersampling pattern.

\begin{figure}[t]
\centerline{\includegraphics[width=1.0\columnwidth]{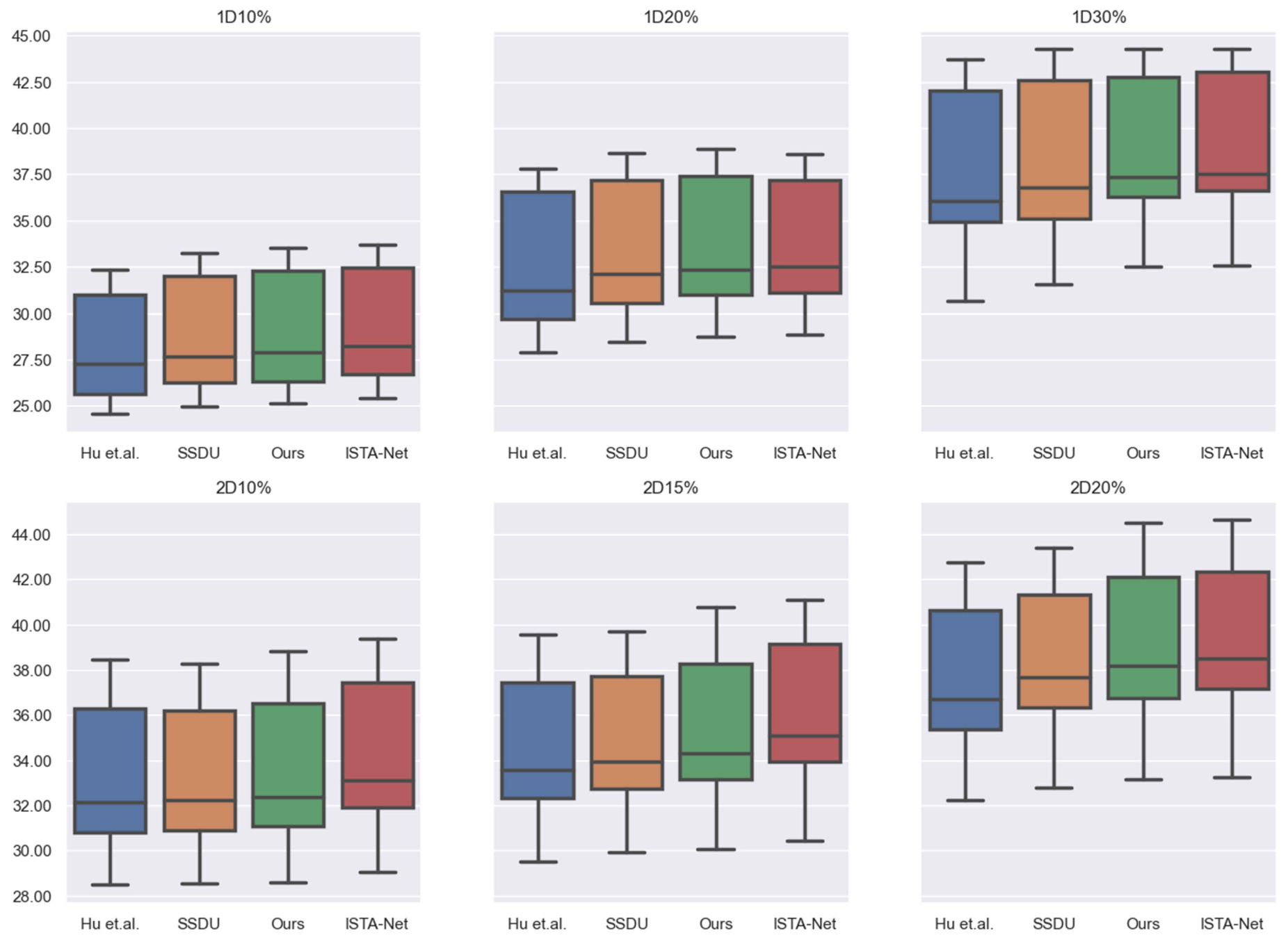}}
\caption{The PSNR results on the single-coil brain MRI dataset.}
\label{fig3}
\end{figure}

\begin{figure}[t]
\centerline{\includegraphics[width=1.0\columnwidth]{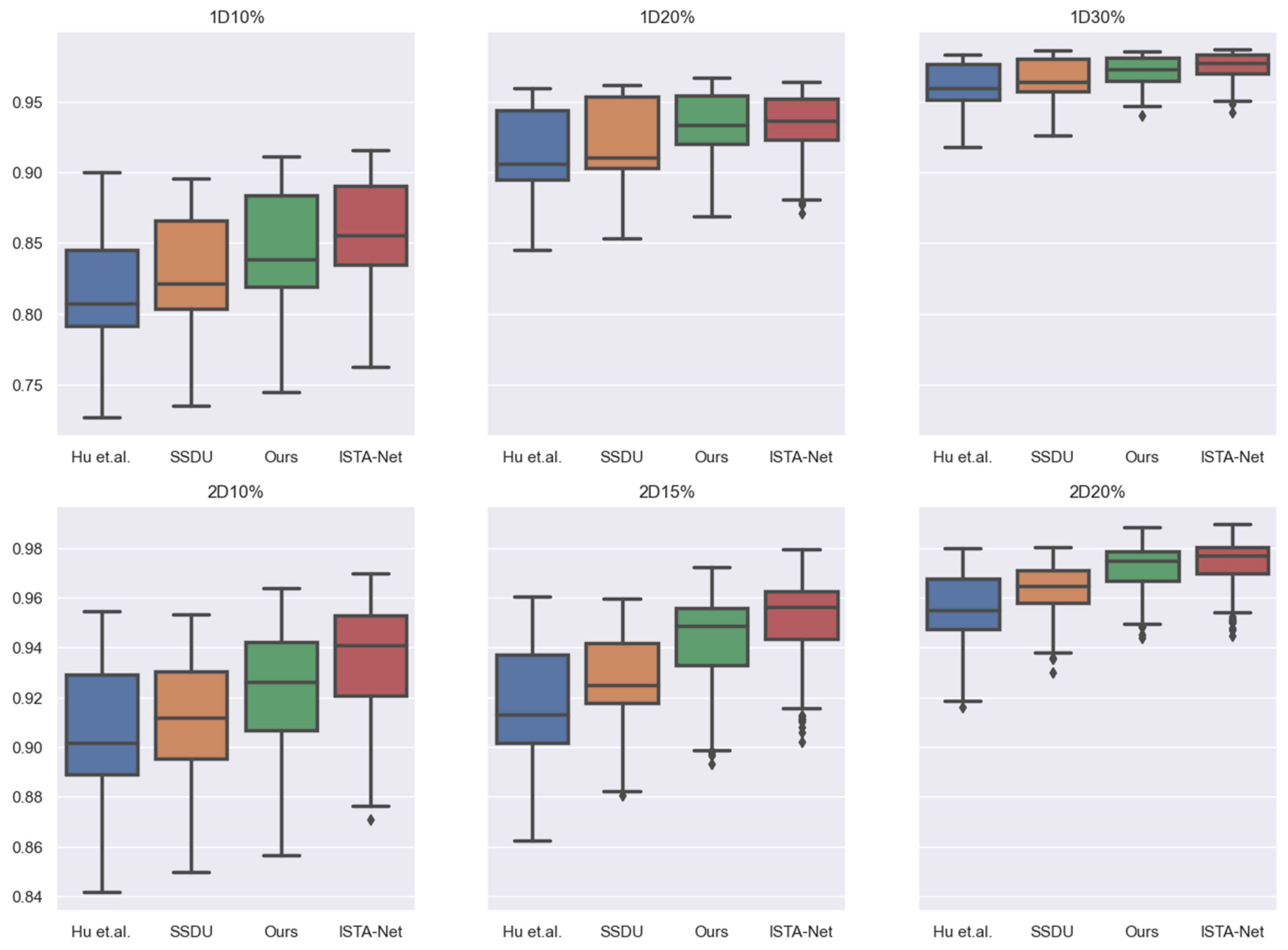}}
\caption{The SSIM results on the single-coil brain MRI dataset.}
\label{fig4}
\end{figure}

\subsection{Image Quality Assessment}
We use Peak Signal-to-Noise Ratio (PSNR in dB) and Structure Similarity Index Measure (SSIM) to evaluate image reconstruction quality. The PSNR index provides voxel-wise reconstruction accuracy calculated as
\begin{equation}\label{eq14}
{\rm PSNR} = 20{\log _{10}}\left( {\frac{1}{{{{\left\| {{I_{\rm pred}} - {I_{\rm gt}}} \right\|}_2}}}} \right),
\end{equation}
where $I_{\rm pred}$ represents the predicted reconstructed image and $I_{\rm gt}$ represents the ground-truth image.

\begin{figure*}[t]
\centerline{\includegraphics[width=0.73\textwidth]{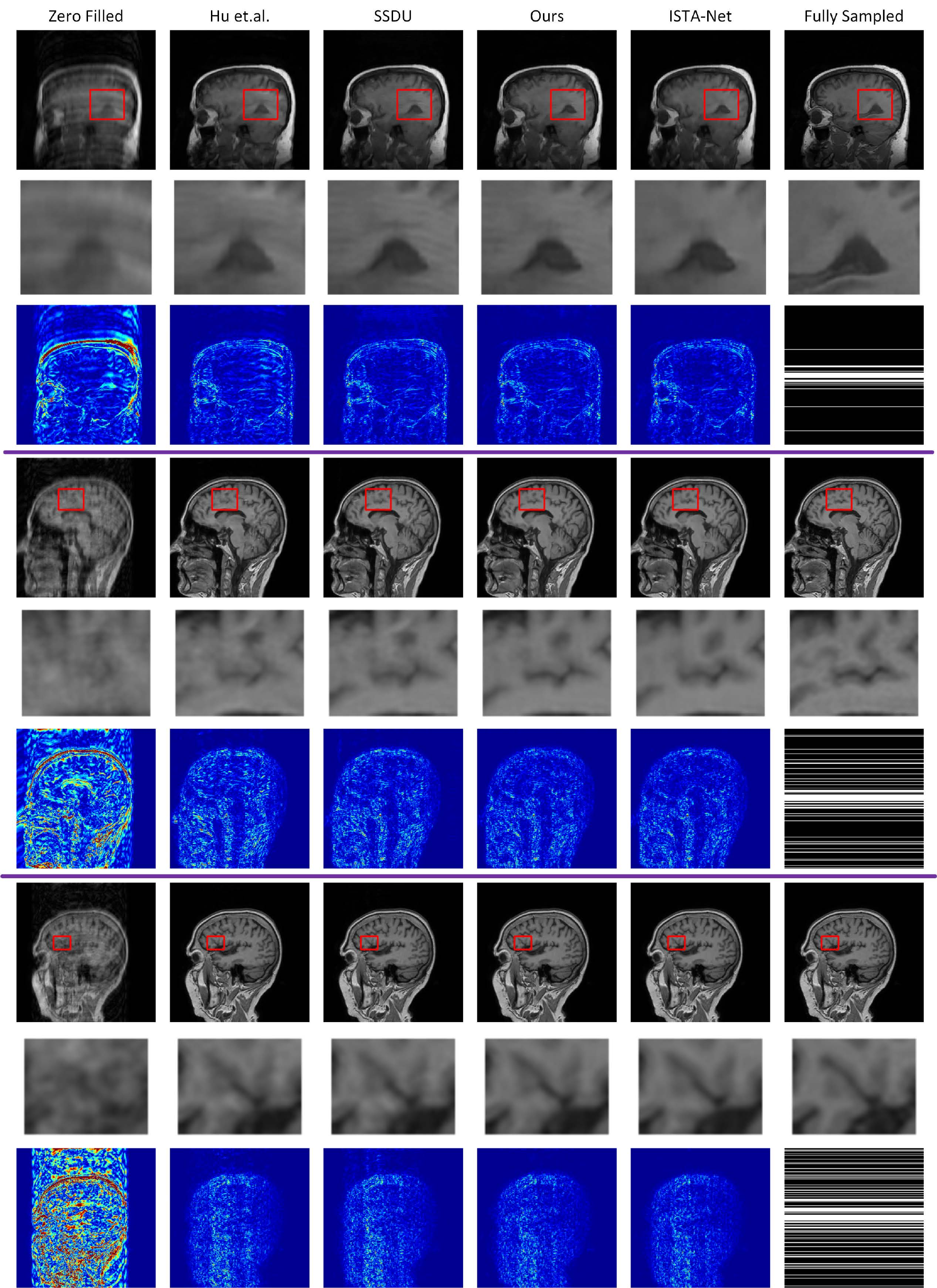}}  
\caption{The subjective reconstruction results of 1D line-based Cartesian undersampling patterns including 1D10$\%$, 1D20$\%$ and 1D30$\%$ from top to bottom.}
\label{fig5}
\end{figure*}

\begin{figure*}[t]
\centerline{\includegraphics[width=0.76\textwidth]{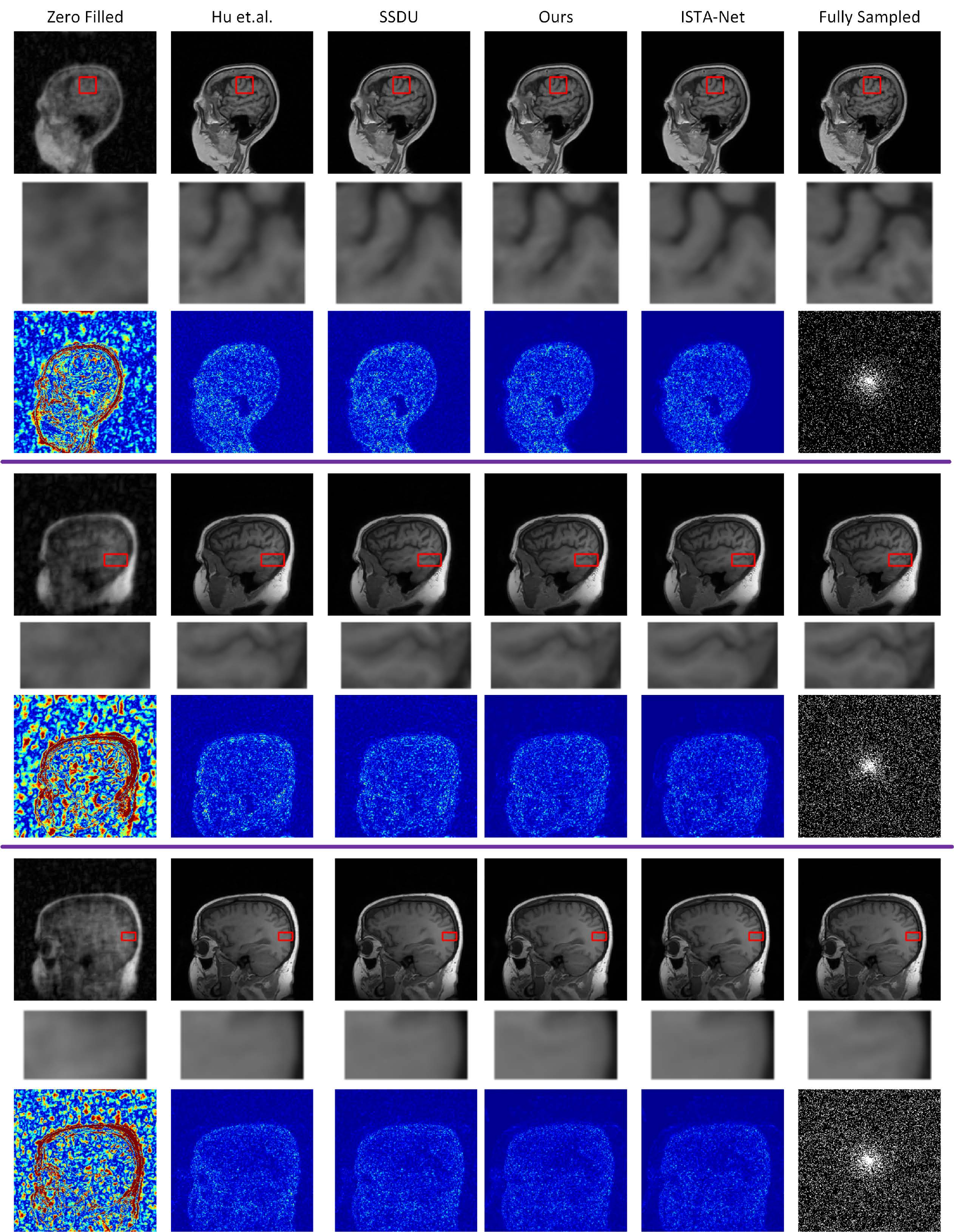}}
\caption{The subjective reconstruction results of 2D Random undersampling patterns including 2D10$\%$, 2D15$\%$ and 2D20$\%$ from top to bottom.}
\label{fig6}
\end{figure*}
The SSIM index evaluate the similarity between the predicted reconstructed image $I_{\rm pred}$ and ground-truth image $I_{\rm gt}$ from the perspectives of luminance, contrast, and structure. Thus it served as a supplementary index. The SSIM index is calculated as
\begin{equation}\label{eq15}
{\rm SSIM} = \frac{{\left( {2{\mu _{{\rm{pred}}}}{\mu _{{\rm{gt}}}} + {c_1}} \right)\left( {{\sigma _{{\rm{pred,gt}}}} + {c_2}} \right)}}{{\left( {\mu _{{\rm{pred}}}^2 + \mu _{{\rm{gt}}}^2 + {c_1}} \right)\left( {\sigma _{{\rm{pred}}}^2 + \sigma _{{\rm{gt}}}^2 + {c_2}} \right)}},
\end{equation}
where  $\mu_{\rm pred}$ and $\mu_{\rm gt}$ represents the mean value of the predicted and ground-truth images, and $\sigma_{\rm pred}$ and $\sigma_{\rm gt}$ represents the corresponding standard deviations. The $\sigma_{\rm pred,gt}$ is the covariance. The notations $c_1$ and $c_2$ are constants.

\renewcommand{\arraystretch}{1} 
\begin{table*}[!tp]
	\centering
	\begin{threeparttable}
		\caption{Performance comparison of multi-coil brain MRI reconstruction in PSNR and SSIM.}
		\label{tab:performance_comparison}
		\begin{tabular}{ccccccc}
			\toprule
			\multirow{2}{*}{Method}&
			\multicolumn{2}{c}{4-Fold}&\multicolumn{2}{c}{6-Fold}&\multicolumn{2}{c}{8-Fold}\cr
			\cmidrule(lr){2-3} \cmidrule(lr){4-5}  \cmidrule(lr){6-7}
			&PSNR&SSIM&PSNR&SSIM&PSNR&SSIM\cr
			\midrule
			Zero Filled   &  30.97 ± 3.84 & 0.863 ± 0.113 & 28.87 ± 3.84 & 0.828 ± 0.126 &  26.48 ± 3.79 & 0.794 ± 0.131 \cr
			CG-SENSE & 32.62 ± 4.73 & 0.851 ± 0.139 & 30.38 ± 4.80 & 0.819 ± 0.144 & 28.72 ± 4.82 & 0.800 ± 0.145 \cr
			SSDU         & 36.08 ± 4.68 & 0.907 ± 0.124 & 32.56 ± 4.46 & 0.871 ± 0.131 & 30.17 ± 4.42 & 0.834 ± 0.132 \cr
			Ours           & \textbf{36.68 ± 4.75} & \textbf{0.917 ± 0.126} & \textbf{33.12 ± 4.67} & \textbf{0.881 ± 0.133} & \textbf{30.70 ± 4.37} & \textbf{0.852 ± 0.132} \cr
			VS-Net       & \textcolor{red}{37.69 ± 4.80} & \textcolor{red}{0.930 ± 0.128} & \textcolor{red}{33.87 ± 4.47} & \textcolor{red}{0.895 ± 0.134} & \textcolor{red}{31.46 ± 4.30} & \textcolor{red}{0.868 ± 0.134} \cr
			\bottomrule
		\end{tabular}
	\end{threeparttable}
\end{table*}

\begin{figure*}[t]
\centerline{\includegraphics[width=0.76\textwidth]{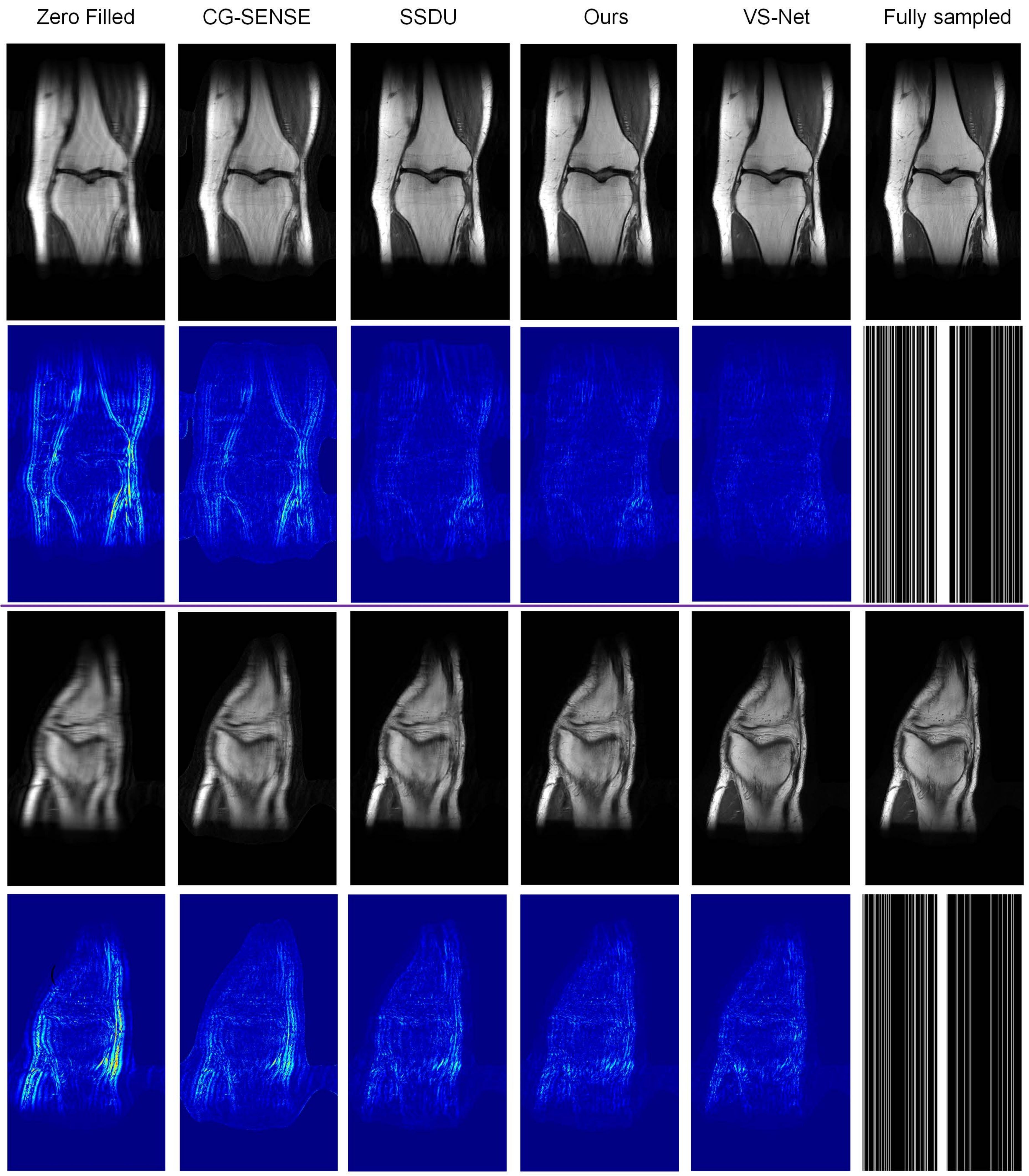}}
\caption{The subjective reconstruction results on multi-coil knee MRI of 1D line-based Cartesian undersampling patterns with acceleration factor of 4-fold and 6-fold from top to bottom.}
\label{fig7}
\end{figure*}

\subsection{Results on Single-Coil Brain MRI Dataset}
We compare the proposed SiamRecon with other two state-of-the-art self-supervised learning approaches: Parallel Networks proposed by Hu et.al. \cite{hu2021self} and SSDU \cite{yaman2020self}. The fully supervised backbone ISTA-Net serves as the upper bound for the self-supervised methods. Specifically, we perform model comparison on the 1D Cartesian undersampling pattern with sampling rate $10\%$, $20\%$ and $30\%$, and 2D Random undersampling pattern with sampling rate $10\%$, $15\%$ and $20\%$.

The PSNR results in box plots are reported in Figure \ref{fig3}. It is observed the proposed SiamRecon offers the closest model performance to the upperbound fully supervised ISTA-Net. The SiamRecon outperforms the SSDU on all the undersampling patterns. Specifically, the PSNR value of reconstruction results of SiamRecon is above 0.5dB higher in PSNR than the one of SSDU on the 20$\%$ 2D sampling pattern. The performance gain is above 0.6dB on the 30$\%$ 1D sampling pattern. The SSDU method achieves higher reconstruction accuracy compared with the Parallel Networks. The SSIM results in box plots are also reported in Figure \ref{fig4}. We observe the compared SSIM results are consistent with the compared PSNR results, which means our SiamRecon achieves both higher voxel-wise reconstruction accuracy and better visual quality.

We show three reconstruction examples in Figure \ref{fig5} on the 1D Cartesian undersampling patterns. For better visual examination, we give local magnification and error maps together with original reconstruction images. Form the results of $10\%$ 1D sampling pattern, the SiamRecon provides sharper and clearer boundaries of cerebrospinal fluid and white matter compared with Parallel Networks, and less undersampling artifact residuals compared with the SSDU. In the case of $20\%$ 1D sampling pattern, the undersampling artifacts are removed for the Parallel Networks, the SSDU and the SiamRecon, however, the fine structure displayed in the magnification box is better preserved in the SiamRecon. Accordingly, the contrast between cerebrospinal fluid and gray matter is well presented. Similarly, low-contrast brain structures are better recovered for our SiamRecon in the case of $30\%$ 1D sampling pattern. The error maps further validate the high-quality image reconstruction of the SiamRecon. We show the corresponding undersampling patterns in the rightmost column.

Another three reconstruction examples are given in Figure \ref{fig6} on the 2D Random undersampling patterns. The MRI reconstruction under 2D sampling patterns is more accurate than the one under 1D Cartesian patterns because of lower coherence, but 1D line-based Cartesian sampling patterns are more easily to realize physically and applicable in practice. The cerebrospinal fluid region is precisely recovered for SiamRecon compared with Parallel Networks and the SSDU on the $10\%$ 2D Random undersampling patterns as shown in the magnification boxes. In the case of $15\%$ 2D Random sampling pattern, We observe the SiamRecon provides closer reconstruction to the fully sampled reference. As demonstrated by the ultra low-contrast case in the reconstruction under $20\%$ 2D pattern, the SiamRecon better recovers this content in the red box compared with other comparative methods even including the fully supervised ISTA-Net.

\begin{figure}[t]
\centerline{\includegraphics[width=\columnwidth]{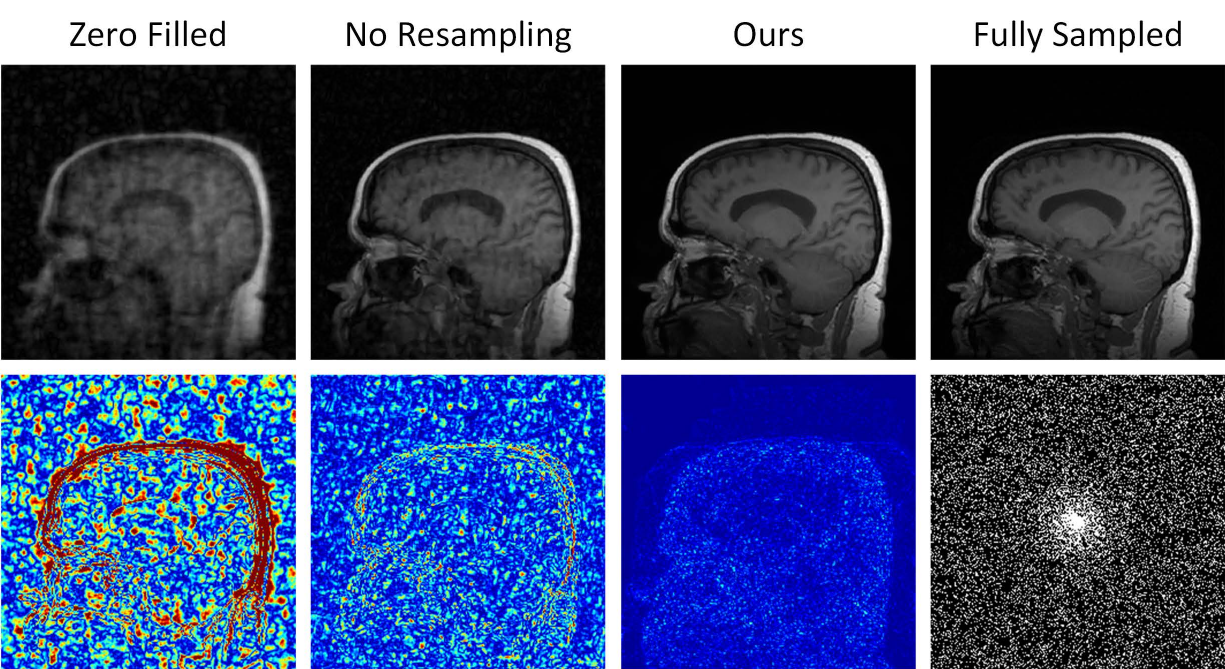}}
\caption{The subjective results of the SiamRecon with and without k-space resampling.}
\label{fig8}
\end{figure}

\begin{figure}[t]
\centerline{\includegraphics[width=\columnwidth]{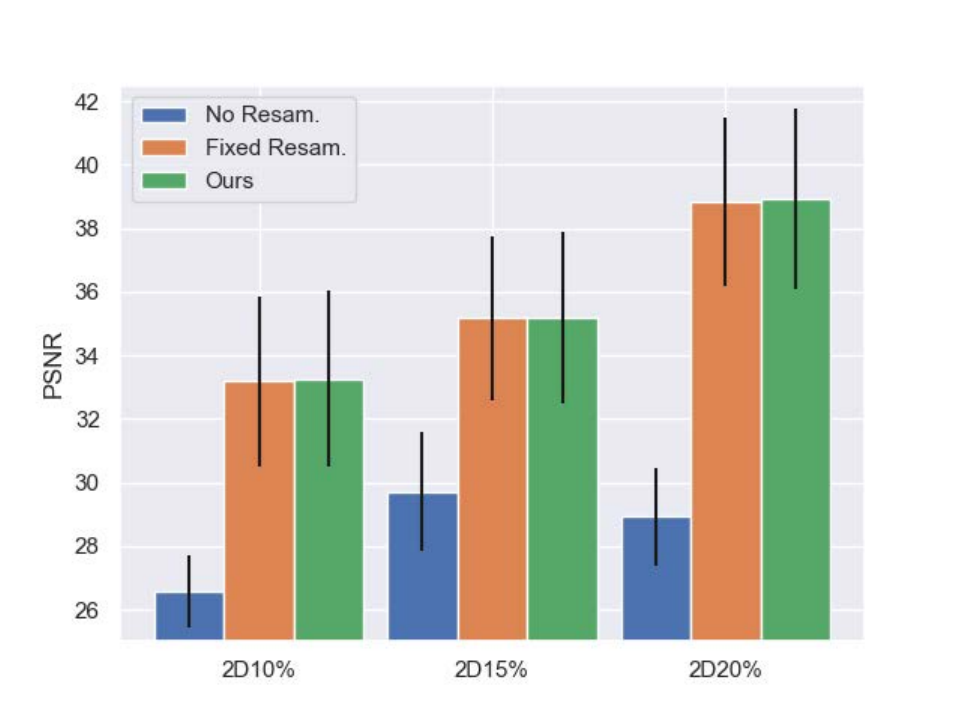}}
\caption{The PSNR results of the SiamRecon with varying/fixed k-spae resampling patterns and the one without resampling.}
\label{fig9}
\end{figure}

\subsection{Results on Multi-Coil Knee MRI Dataset}
We also evaluate the proposed SiamRecon on the multi-coil knee MRI dataset. The comparison is performed with alternate methods including the CG-SENSE \cite{pruessmann1999sense}, the SSDU \cite{yaman2020self} and fully supervised VS-Net \cite{duan2019vs}. The 1D Cartesian undersampling patterns are used with acceleration factors including 4 folds, 6 folds and 8 folds. We report the averaged PSNR and SSIM results in Table \ref{tab:performance_comparison}. Similar to the experimental results for single-coil MRI, the SiamRecon achieves better reconstruction accuracy compared with SSDU and approximates the VS-Net under full supervision. Two reconstruction examples are given in Figure \ref{fig7} with 4-fold and 6-fold acceleration, respectively. The SiamRecon suffer from less undersampling artifacts compared with the SSDU. We observe the SiamRecon shows minimal reconstruction errors except the fully supervised VS-Net according to the residual error maps.

\begin{figure}[h]
\centerline{\includegraphics[width=\columnwidth]{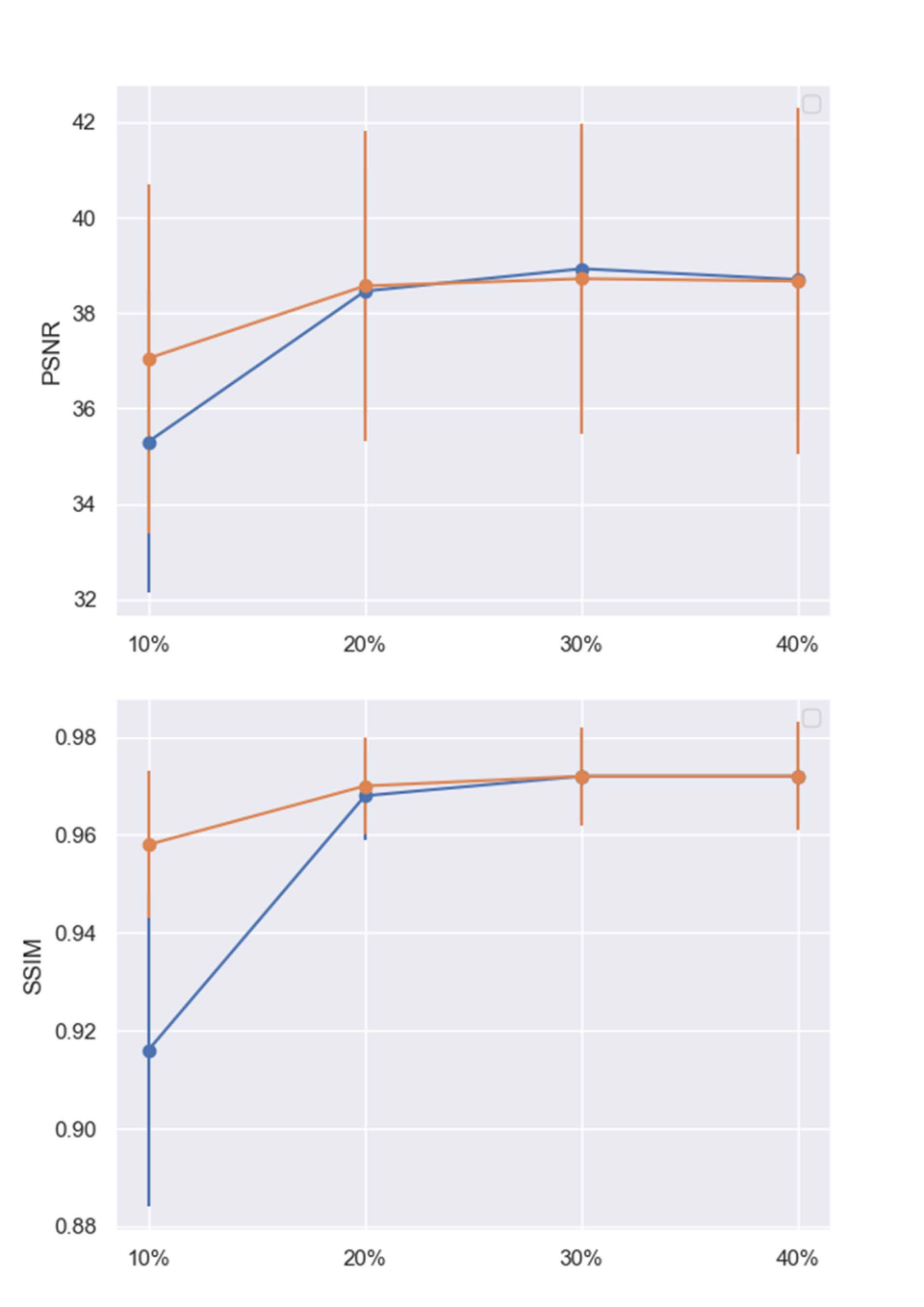}}
\caption{The PSNR and SSIM results of the SiamRecon with different k-space resampling ratios.}
\label{fig10}
\end{figure}

\renewcommand{\arraystretch}{1} 
\begin{table*}[tp]
	\centering
	\begin{threeparttable}
		\caption{The experimental results of the ablation study.}
		\label{tab:ablation_table}
		\begin{tabular}{cccccccccc}
			\toprule
			\multirow{2}{*}{Method}&
			\multirow{2}{*}{K-Space}&\multirow{2}{*}{Stop}&\multirow{2}{*}{Parameter}&\multicolumn{2}{c}{2D10$\%$}&\multicolumn{2}{c}{2D15$\%$}&\multicolumn{2}{c}{2D20$\%$} \cr
			\cmidrule(lr){5-6} \cmidrule(lr){7-8} \cmidrule(lr){9-10}
			&Resampling&Gradients&Replacement&PSNR&SSIM&PSNR&SSIM&PSNR&SSIM\cr
			\midrule
			\textcolor{black}{Ours}   &  \Checkmark & \Checkmark & \Checkmark & \textbf{33.26 ± 2.77} & \textbf{0.923 ± 0.023} & \textbf{35.20 ± 2.69} & \textbf{0.943 ± 0.017} & \textbf{38.93 ± 2.83} & \textbf{0.972 ± 0.008} \cr
			\textcolor{black}{Model${_1}$} & \Checkmark & \Checkmark & \XSolid & 33.18 ± 2.80 & 0.921 ± 0.023 & 35.14 ± 2.63 & 0.941 ± 0.017 & 38.84 ± 2.96 & 0.971 ± 0.010  \cr
			\textcolor{black}{Model${_2}$} & \Checkmark & \XSolid &\Checkmark & 33.00 ± 2.66 & 0.920 ± 0.023 & 35.05 ± 2.78 & 0.943 ± 0.019 & 38.69 ± 2.65 & 0.971 ± 0.010 \cr
			\bottomrule
		\end{tabular}
	\end{threeparttable}
\end{table*}

\subsection{Effects of K-Space Resampling}
\subsubsection{Removing Resampling}
The k-space resampling strategy is a key component in our SiamRecon approach. Suppose the k-space resampling is removed, the M-step in Equation \ref{eq8} turns into the Equation \ref{eq71}. The optimization of Equation \ref{eq71} would yield trivial solution by adapting the network with parameter $f_{\theta}$ to be an identical mapping. We compare the reconstruction results of the original SiamRecon and the one with k-space resampling removed. Without k-space resampling, the SiamRecon fails to remove undersampling artifacts as shown in Figure \ref{fig8}. We further experiment with SiamRecon with fixed k-space resampling mask and give the PSNR comparison in Figure \ref{fig9}. We observe the SiamRecon with varying k-space resampling mask slightly outperforms the one with fixed mask. Both the two versions of SiamRecon achieve significantly better results compared with the one without k-space resampling, demonstrating the SiamNet is robust to the variation of the k-space resampling mask.

\subsubsection{Resampling Rate}
In this section, we test k-space resampling masks with various sampling rates, and report the PSNR and SSIM results in Figure \ref{fig10}. According to this results, we set the sampling rate of the k-space resampling mask to 30$\%$ throughout this work.

\subsection{Ablation Study}
In this section, we validate the parameter replacement and stop gradient in the SiamRecon. The parameter replacement and stop gradient operation are ablated from the SiamRecon, respectively, and we have the corresponding Model$_1$ and Model$_2$. The stop gradient is inspired by the SimSiam approach and the parameter replacement is inspired by the EM algorithm. The stop gradient is proven to help generate stable supervision. The parameter replacement is implemented as a E-step in the EM framework. The comparison results are reported in Table \ref{tab:ablation_table}. We observe the regular SiamRecon mimicking the EM algorithm outperforms the one without stop gradient and parameter replacement, demonstrating the effectiveness to formulate self-supervised MRI reconstruction into a EM framework.

\section{Conclusion}
In this paper, we propose a Siamese-based self-supervised approach for fast MRI reconstruction named SiamRecon, which is motivated by a expectation maximization perspective. The experimental results show the SiamRecon method outperforms the existing self-supervised MRI reconstruction approaches. The proposed SiamRecon provides a promising approach to be applied in clinical MRI practice where moving subject makes it difficult to acquire high-quality full-sampled MRI training references or specific imaging protocal such as EPI. In the future work, we plan to extend the SiamRecon to dynamic MRI reconstruction and MR parameter mapping.


\bibliographystyle{IEEEtran}

\bibliography{S4}

\vfill

\end{document}